\def\draftversion{false}

\RequirePackage{ifthen}
\ifthenelse{\equal{\draftversion}{true}}{
	\documentclass[aps,prb,10pt,galley,amsmath,amssymb,
	floatfix,
	citeautoscript]{revtex4}
}{
	\documentclass[aps,prb,10pt,twocolumn,amsmath,amssymb,longbibliography,
	citeautoscript]{revtex4-1}
}

\usepackage{graphicx}
\usepackage[usenames,dvipsnames]{color} 
\usepackage{bm} 
\usepackage[update,prepend]{epstopdf}
\usepackage{soul} 
\usepackage{dcolumn} 
\usepackage{multirow}
\usepackage{tabularx}
\usepackage{amssymb}

\newcolumntype{C}[1]{>{\hsize=#1\hsize\centering\arraybackslash}X}%

\def\comment#1{}

\ifthenelse{\equal{\draftversion}{true}}{
	\marginparwidth 2.7in
	\marginparsep 0.5in
	\newcounter{comm} 
	\def\commnext{\stepcounter{comm}}
	\def\commtext{{\bf\color{blue}[\arabic{comm}]}}
	\def\commmar{{\bf\color{blue}[\arabic{comm}]}}
	\def\dvm#1{\commnext\marginpar{\small DV\commmar: #1}\commtext}
	\def\jkm#1{\commnext\marginpar{\small JK\commmar: #1}\commtext}
	\def\mlab#1{\marginpar{\small\bf #1}}
	
}{
	\def\dvm#1{}
	\def\jkm#1{}
	\def\mlab#1{}
	
}

\begin{document}

\title{Bismuth antiphase domain wall: A three-dimensional manifestation of the Su-Schrieffer-Heeger model}

\author{Jinwoong Kim$^1$}
\author{Cheng-Yi Huang$^{1,2}$}
\author{Hsin Lin$^3$}
\author{David Vanderbilt$^4$}
\author{Nicholas Kioussis$^1$}

\affiliation{$^1$
	Department of Physics and Astronomy,
	California State University,
	Northridge, California 91330, USA
}
\affiliation{$^2$
	Department of Physics,
	Northeastern University,
	Boston, Massachusetts 02115, USA
}
\affiliation{$^3$
	Institute of Physics,
	Academia Sinica,
	Taipei 11529, Taiwan
}
\affiliation{$^4$
	Department of Physics and Astronomy,
	Rutgers University,
	Piscataway, New Jersey 08854-8019, USA
}

\date{\today}
\begin{abstract}
The Su, Schrieffer and Heeger (SSH) model, describing the
soliton excitations in polyacetylene due to the formation of
antiphase domain walls (DW) from the alternating bond pattern,
has served as a paradigmatic example of one-dimensional (1D)
chiral topological insulators.
While the SSH model has been realized in photonic and plasmonic
systems, there have been limited analogues in three-dimensional
(3D) electronic systems, especially regarding the formation of
antiphase DWs.
Here, we propose that pristine bulk Bi,
in which the dimerization of (111) atomic layers renders
alternating covalent and van der Waals bonding
within and between successive  $(111)$
bilayers, respectively, serves as a 3D analogue of the SSH model.
First, we confirm that the two dimerized Bi structures belong to
different Zak phases of 0 and $\pi$ by considering the parity
eigenvalues and Wannier charge centers, while the previously reported bulk topological phases of Bi remain invariant
under the dimerization reversal.
Next, we demonstrate the existence of
topologically non-trivial (111) and trivial
$(11\bar{2})$ DWs in which the number of in-gap DW states
(ignoring spin) is odd and even respectively, and show how this
controls the interlinking of the Zak phases of the two adjacent domains.
Finally,  we derive general criteria specifying when a DW of arbitrary
orientation exhibits a $\pi$ Zak phase based on the flip
of parity eigenvalues.
An experimental realization of dimerization in Bi and
the formation of DWs may be achieved via intense femtosecond laser
excitations that can alter the interatomic forces and bond lengths.
\end{abstract}

\maketitle

\section{Introduction}
\label{sec:intro}
\begin{figure*}[t]
	\centering
	\includegraphics[width=17.0cm]{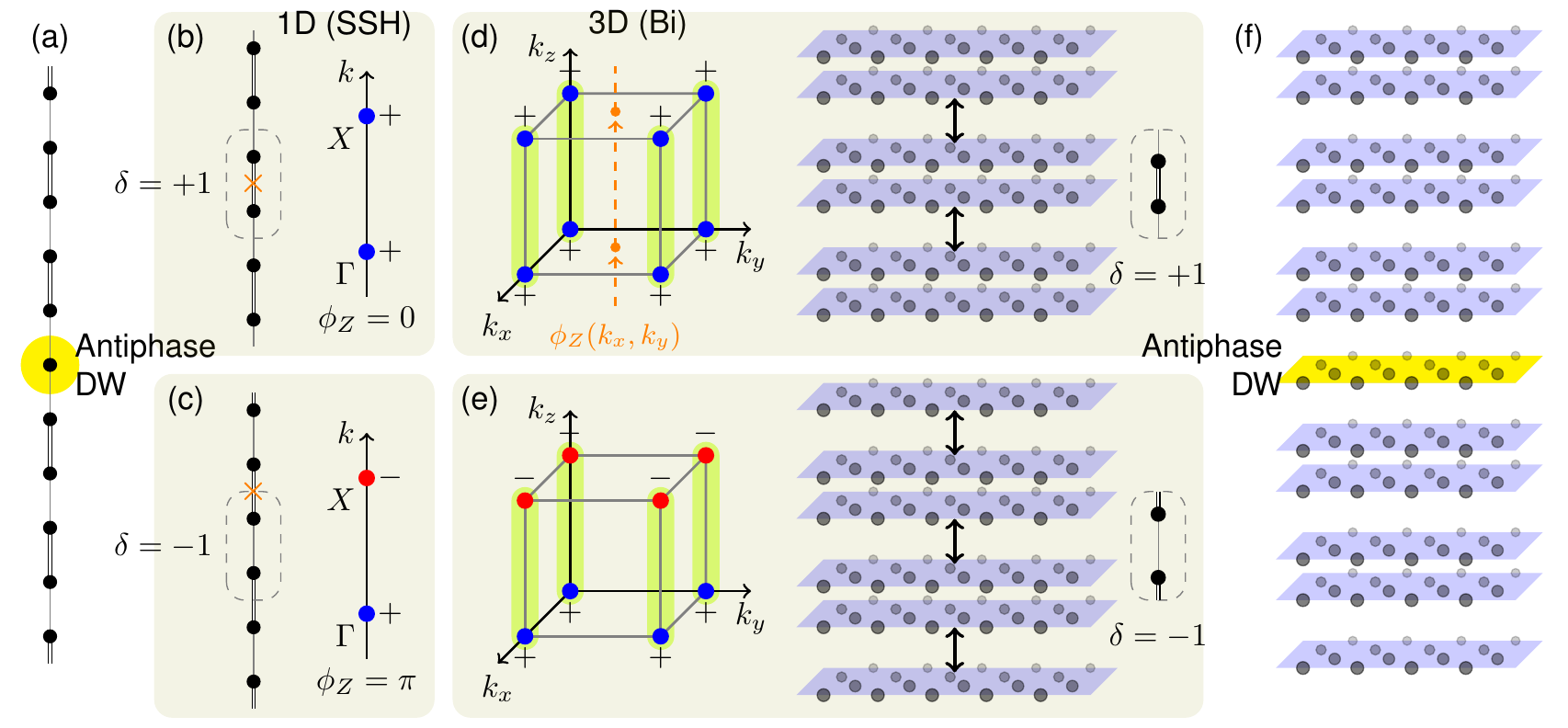}
	\caption{Schematic view of the SSH model and comparison with the 3D analogue. (a) Antiphase DW of the SSH model where two atomic chains with alternating weak and strong bonding are connected out of phase. (b) and (c) Two dimerized phases ($\delta=\pm1$) and corresponding parity eigenvalues at the two time-reversal invariant momenta (TRIM), $\Gamma$ and $X$, whose product determines the Zak phase ($\phi_Z=0,\pi$). Dashed box denotes the repeating unit cell where weak (strong) bonds are trimmed at the cell boundary for $\delta=+1$ ($\delta=-1$). Red crosses indicate the (net) Wannier charge center $\bar{r}$.
	(d) and (e) Parity eigenvalues at the eight TRIM points of the 3D SSH model for the two dimerized $\delta = \pm1$ phases shown on the right, with strong (weak) intra- (inter-) bilayer bonding. The shaded vertical lines in momentum space correspond to four individual 1D SSH models in the presence of inversion and time-reversal symmetries whose Zak phase is quantized and flipped by the dimerization reversal in analogy to the SSH model. 
	Vertical dashed line in (d) illustrates the closed 1D path, $k_z \in [0,2\pi]$ along which the Zak phase is defined.
	(f) Antiphase DW as the 3D analogue of the SSH model where the Zak-phase-induced in-gap states emerge at the four interface TRIM points, spatially localized at the central (yellow) layer.
	\label{fig:schematic}
	}
\end{figure*}

Polyacetylene,
\cite{Heeger_NobelLecture}
(CH)$_x$, is an infinite one-dimensional (1D) carbon
chain whose {\it trans} configuration has two degenerate
dimerized structures consisting of alternating double and single
bonds which can be interchanged by symmetry.
Interestingly, polyacetylene exhibits finite electric conductivity
even though its intrinsic band structure is insulating.
This can be understood in terms of the migration of electrically-charged
antiphase domain walls (DWs) between two structures (domains) with
opposite dimerization 
as illustrated in Fig.~\ref{fig:schematic} (a-c).
The Su-Schrieffer-Heeger (SSH) model,
\cite{SSH_PRL,SSH_PRB}
introduced to describe polyacetylene, yields a transition
from a trivial to topological non-trivial phase depending on
the relative hopping amplitudes between the two distinct types
of bondings, where the so-called ``winding number" undergoes a
discontinuous change from $0 \rightarrow 1$.
The ``winding number" is closely related to the Zak phase
\cite{Zak_PRL1989}
which is quantized to be 0 or $\pi$ for systems with space inversion symmetry.
Moreover, the DW in the SSH model leads to the emergence
of a 
boundary localized zero-energy mode
in the middle of the energy gap with charge accumulation of
$\pm e/2$, analogous to the fractionally charged
excitations in quantum field theory.
This midgap state is understood as a topologically protected
boundary mode and the SSH model serves as a paradigmatic example
of topological insulator protected by a chiral (i.e., sublattice) symmetry.

The SSH model is the simplest and one of the most important models
in describing band topology in condensed matter physics, and
has been the subject of intense investigations such as
Majorana zero mode in a finite atomic chain
\cite{Kitaev_Majorana_2001,Yu_Majorana_2020}
and an
extension to two-dimensional (2D) systems, including
graphene\cite{Delplace_PRB_2011} and  four-basis-\cite{Liu_PRL_2017,Obana_PRB2019} and
two-basis-\cite{Prodan_PRB_2019} square-lattice models.
The latter study\cite{Prodan_PRB_2019}
explicitly characterized several topological phases with distinct winding numbers upon uniaxial strain and sublattice dimerization where zero-energy flat bands were predicted to emerge on 1D antiphase DW if the winding numbers (equivalently the Zak phases) of the two facing domains are different.
This is the 2D analogue of the SSH model.

In three-dimensional (3D) systems, non-trivial $\pi$ Zak phases have drawn less attention and only a few systems have been found to exhibit them.
\cite{Murakami_PRX_2018,Murakami_PRR_2020}
Sc$_2$C, a designed inorganic electride,
\cite{Zhang_PRX} 
was predicted to exhibit a
$\pi$ Zak phase with consequent surface states inside its insulating
band gap.\cite{Murakami_PRX_2018}
Surface drum-head states of topological nodal-line semimetals
are also known to originate from the $\pi$ Zak phase,
\cite{Ryu_Nodal_2002,Burkov_Nodal_2011,Fang_2016}
where the drum-head states are bounded by surface-projected
bulk nodal lines, in contrast to the $\pi$ Zak phase insulator.
For example, Sc$_2$C (Y$_2$C) is a $\pi$ Zak phase insulator (topological nodal-line semimetal) where
the surface states cover 100\% (90.4\%) of the surface BZ.
\cite{Murakami_PRX_2018}
The $\{111\}$
surfaces of silicon and diamond 
host surface bands from the $\pi$ Zak phase,
\cite{Murakami_PRR_2020} 
where each surface unit cell accumulates one half of an electron
\cite{Vanderbilt}
leading to half-filled metallic surface bands. 
An insulating surface can be achieved only by 
even-number (such as $2 \times 2$)
surface reconstructions that allow an integer number of surface 
electrons and hence fully filled bands.
\cite{Murakami_PRR_2020,Smeu_PRB}
The 3D $\pi$ Zak phase systems that have been reported so far involve no atomic displacement that can be described as a dimerization.
In this work, we show that each (111) atomic layer of Bi corresponds to a single site of the 1D SSH model, and dimerization of the atomic layers in the ground state results in $0$ or $\pi$ Zak phases depending on the dimerization sign, 
$\delta=\pm1$, as illustrated in Fig.~\ref{fig:schematic}(d-e).

\begin{figure*}
	\centering
	\includegraphics[width=17.0cm]{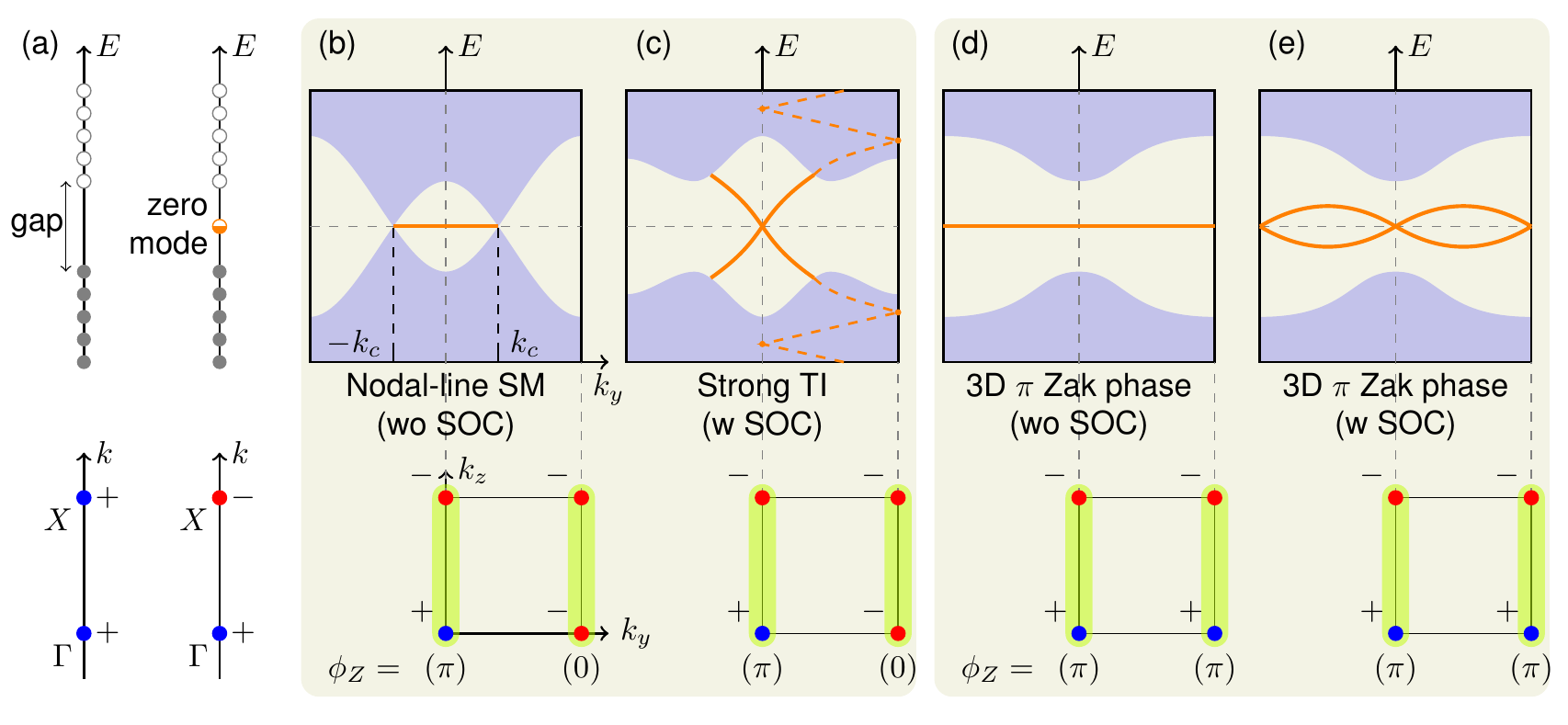}
	\caption{Schematic view of the Zak phase, $\phi_Z (k_x=0,k_y)$
	and relevant boundary states in several topological systems. Lower (upper) panels illustrate the parity eigenvalues at TRIM points (boundary band structure). 
	(a) 1D SSH model. Open (closed) circles denote empty (filled) states of the zero-dimensional boundary. Non-trivial $\pi$ Zak phase on the right side induces a half-filled zero mode at the Fermi level. 
	(b-e) $k_x = 0$ plane of the 3D momentum space and its surface (normal to $z$) band structure without and with SOC. Orange solid lines (shaded areas) denote surface (surface-projected bulk) states.
	(b) Nodal-line semimetal whose quantized Zak phase is $\pi$ and $0$ for $|k_y| < |k_c|$ and $|k_y| > |k_c|$, respectively. The continuous zero mode at $|k_y| < |k_c|$ forms drum-head states, connecting boundary-projected bulk nodal lines.
	(c) Strong topological insulator with a surface Dirac cone which emerges at a surface TRIM with $\pi$ Zak phase. Dashed orange lines illustrate surface band connectivity between surface TRIM, referred to as ``switch partners". (d) and (e) 3D $\pi$ Zak phase without and with SOC, respectively. 
	In the presence of SOC, the Zak phase is no longer quantized except at the surface TRIM and the in-gap state splits at generic momentum $k$.
	\label{fig:schematic_b}	
	}
\end{figure*}

The Zak phase
\cite{Zak_PRL1989}
is a special form of the Berry phase
\cite{BerryPhase}
and is equivalent to the electronic part of the polarization,
\cite{King-Smith,Vanderbilt,Resta1994}
\begin{eqnarray}
	\label{eq:Zak}
	\phi_Z = \phi_B = \frac{2\pi p}{e c},
\end{eqnarray}
where $-e$ is the electron charge, $c$ is the lattice constant
of the unit cell,
and $p$ is the dipole moment of the bulk unit cell
that can be in turn expressed in terms of the Wannier functions
of the occupied bands,
\begin{eqnarray}
	\label{eq:dipole}
	p = -\sum_{i}^{\textrm{occ.}} e r_i.
\end{eqnarray}
Here, $r_i$ is the center of $i$-th Wannier function.
In the presence of inversion symmetry, the dipole moment is
quantized such that the Zak phase can only take on values
$\phi_Z=0$ or $\pi$, corresponding to whether the net
Wannier center $\bar{r} = \sum_{i} r_i$ is located at the center
($\bar{r}=0$) or the boundary ($\bar{r}=c/2$) of the bulk unit cell,
respectively
(see Fig.~\ref{fig:schematic}(b,c)).
This definition depends on the choice of inversion
center for the placement of the origin, which we assume to have
been decided once and for all.
The origin-dependent Zak phase has been discussed
in detail by introducing the ``intercellular Zak phase".\cite{Rhim_PRB_2017}
The Zak phase can be easily computed from the product of the parity
eigenvalues of the occupied bands at time-reversal
invariant momenta (TRIM) 
\cite{FuKaneMele_PRL,FuKane_PRB}
in the 1D momentum space, where
$\phi_Z = 0$ ($\pi$) corresponds to positive (negative) product
as shown in Fig.~\ref{fig:schematic} (b)
(Fig.~\ref{fig:schematic} (c)).

In 2D and 3D systems, the Zak phase can be
defined on a closed 1D path such as 
a periodic $k_z$ string with a fixed in-plane momentum ($k_x, k_y$)
as shown in Fig.~\ref{fig:schematic}(d).
Under inversion and time-reversal symmetry and in the absence of
SOC, an insulating bulk has 
a constant and quantized Zak phase on 
such strings normal to a given surface regardless of the
specific in-plane momentum ($k_x, k_y$).
This implies that a single pair of surface-projected TRIM is
enough to determine the Zak phase of the entire surface,
$\phi_Z (k_x,k_y) = \phi_Z (0,0) = \{0,\pi\}$.
Turning on the SOC, however, allows modulation of the Zak phase 
which is no longer quantized at generic surface momenta 
except at the surface TRIM.
Because of the strong SOC of Bi, we focus on the four 
surface TRIM where the Zak phase is quantized to be
0 or $\pi$ (see shaded lines in Fig.~\ref{fig:schematic}(d,e) 
connecting the four pairs of TRIM).

Figure~\ref{fig:schematic_b} illustrates schematically the boundary states of various topological phases and the
corresponding Zak phase configurations at the surface TRIM points. 
Hirayama \textit{et al.}~\cite{Murakami_PRX_2018} 
demonstrated that the surface states of the 3D $\pi$ Zak phase (Fig.~\ref{fig:schematic_b}(d)) is a full-BZ extension of the drum-head states of a nodal-line semimetal (Fig.~\ref{fig:schematic_b}(b)).
This can be understood as a continuous shift of $k_c \rightarrow \pi$ that accompanies a band inversion at $k_y = \pi$ and switching of the Zak phase $\phi_Z (0,\pi)$ from $0$ to $\pi$.
In the presence of SOC, the degeneracy of the bulk nodal lines is lifted 
and the system becomes a strong topological insulator (STI) as shown in Fig.~\ref{fig:schematic_b}(c). Note that the Zak phases 
$\phi_Z(0,0)=\pi$ and $\phi_Z(0,\pi)=0$ do not change.
In contrast to the STI phase where 
a robust surface state is guaranteed by the ``switch partners" 
band connectivity between TRIM,
\cite{FuKane_PRB,FuKaneMele_PRL}
the surface state induced by the $\pi$ Zak phase is rather
isolated in energy from the valence and conduction 
bands (Fig.~\ref{fig:schematic_b}(e)).
The surface band can be pushed into the valence or conduction
bands via surface modifications unless it is protected by
a chiral (or particle-hole) symmetry which in turn pins the non-trivial
surface state at the Fermi level.
Since the Bi $p$ bands are well separated from
the lower energy $s$ bands 
and the inter-sublattice hopping matrix elements
($\sigma_{w,v}$ in Appendix A) are dominant
there is an effective chiral symmetry
which retains the non-trivial state within the 
bandgap of the Bi antiphase DW.

In this work we propose that the $\alpha$-phase of bulk Bi in the
rhombohedral structure is a 3D analogue of the 1D SSH system.
In Sec.~\ref{sec:bulk} using DFT-parameterized tight-binding model
calculations we investigate the topological properties of the two
dimerized states of bulk Bi.
We find that the dimerization reversal induces parity sign flip
at four TRIM
(without changing the bulk topology)
which in turn induce a transition of the Zak phase from
$\pi\rightarrow$ 0, consistent with the emergence of 
odd or even
number of Wannier charge centers (WCCs) at the cell boundary.
In Sec.~\ref{sec:dw_111} we consider two types of (111) DWs sandwiched between
two oppositely dimerized states and show the emergence of
topologically-protected DW localized states,
in contrast to the trivial DW states for the (11$\bar{2}$)
DWs discussed in Sec.~\ref{sec:dw_112}.
In Sec.~\ref{sec:dw_arbitrary}
we derive criteria for the emergence of 
$\pi$ and $0$ Zak phases
for a DW of arbitrary orientation
and identify those DW orientations that host non-trivial 
DW states.
Sec.~\ref{sec:exp_realization}
discusses a plausible experimental realization of the 
dimerization reversal in pristine Bi and the formation of DW 
using intense femtosecond laser excitations that can alter the
interatomic forces and energy barriers 
between the two dimerized states.
\cite{Fritz_Science}
Conclusions are summarized in Sec.~\ref{sec:conclusion} 
and Sec.~\ref{sec:methods} describes the methodology used.
Our findings suggest a novel band engineering concept
for topologically protected states using antiphase
DWs where the parity sign flip can occur without the
assistance of strong spin-orbit coupling of heavy ions.

\setlength{\unitlength}{1cm}
\begin{figure}[t]
	\centering
	\includegraphics[width=8.6cm]{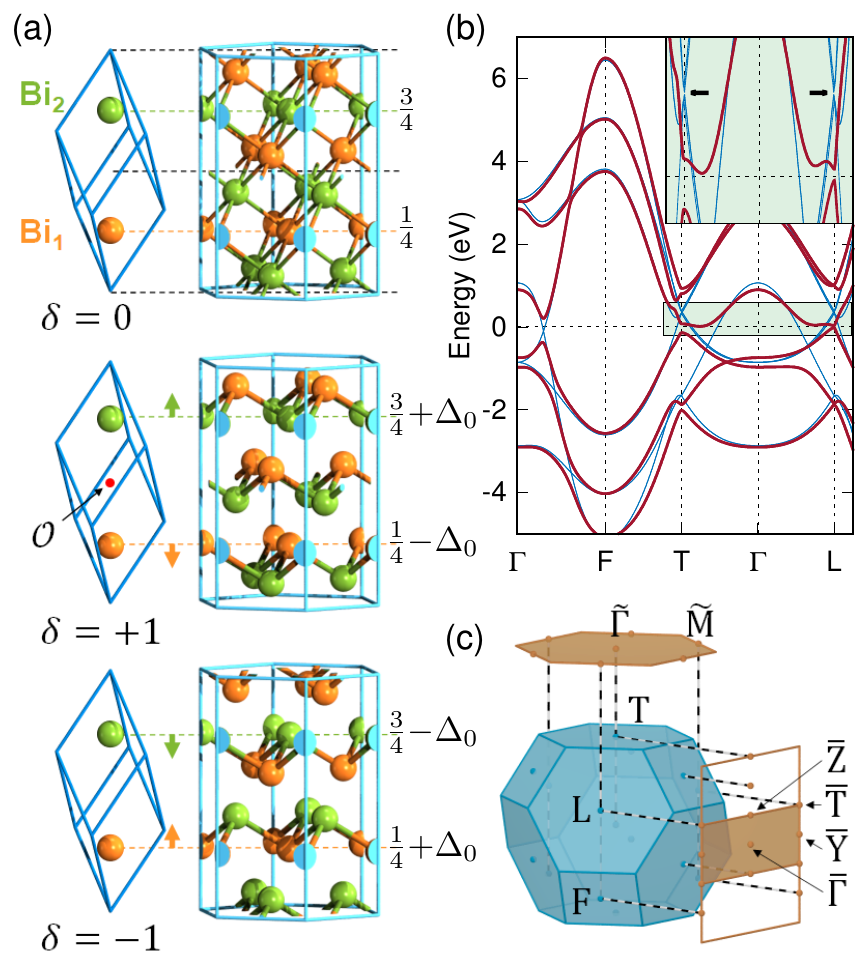}
	\caption{
	(a) The primitive cell of Bi with two sublattice
	sites at fractional height
	$1/2 \pm (1/4 + \Delta)$ along the $[111]$ direction,
	where
	$\Delta$ is the Bi displacement
	$\delta=\Delta/\Delta_0=\pm1$
	is the dimerization sign, and $\Delta_0$ is the equilibrium
	displacement. 
	Top panel denotes the unstable undimerized state ($\delta=0$) and the two lower panels denote the stable dimerized states ($\delta=\pm1$).
	Dimerized Bi atomic layers, stacked along
	$[111]$, are illustrated in the conventional hexagonal cell
	on the right.
	The inversion center $\mathcal{O}$
	located at the center of the primitive cell is marked with the red dot.
	(b) The calculated bulk band structure using
	tight-binding parameters obtained from
	first-principles calculations (see Appendix~\ref{sec:tb}).
	Red (blue) lines denote the stable
	dimerized (unstable undimerized) structure.
	Inset: Zoom-in band structure near the $T$ and $L$ points
	showing the narrow gap and gap-closing (marked by arrows)
	for the dimerized and undimerized structures, respectively.
	(c) First Brillouin zone (BZ) of bulk Bi and its
	projection on the (111) and (11$\bar{2}$)
	interface BZs.
	\label{fig:prim}
	}	
\end{figure}

\section{Bulk Bi: Dimerization and Topology}
\label{sec:bulk}
{\it Atomic structure --} The $\alpha$-phase of bulk Bi
in the rhombohedral structure (space group $R\bar{3}m$, No. $166$)
is shown in Fig.~\ref{fig:prim}(a), where the conventional unit cell
has a bilayer (BL) structure with an ABC stacking sequence along the
$[111]$ direction consisting of three BLs.
There is strong covalent bond within each BL (intra-BL bonding),
with a Bi atom forming three $\sigma$ bonds with its nearest
neighbors, and weak van der Waals bonding between two
nearest-neighbor BLs (inter-BL bonding).
The intra- and inter-BL sequence of bonds alternate along the
$[111]$ stacking direction, which is exactly analogous to the
alternating double and single bonds in polyacetylene shown
schematically in Fig.~\ref{fig:schematic}(b-e).

Furthermore, as shown in Fig.~\ref{fig:prim}(a), 
the intra- and inter-BL bonds
can be interchanged, resulting in two degenerate dimerized
ground states with opposite dimerization parameters
$\delta \equiv \Delta/\Delta_0 = \pm 1$. 
Here, $\Delta$ is the displacement of the two Bi atoms
in the primitive cell along $[111]$ [Fig.~\ref{fig:prim}(a)]
in units of the lattice vector
$c=|\vec{a_1}+\vec{a_2}+\vec{a_3}|$ ($\vec{a_i}$, $i=$1-3
are primitive lattice vectors),
and $\Delta_0$ is the equilibrium displacement.
The positively dimerized state can be obtained from the negatively
dimerized state via a translation by a half lattice vector, or vice
versa.
In sharp contrast to 2D and 3D topological orders, the Zak phase
is not invariant under such a translation.

{\it Electronic Structure--} Fig.~\ref{fig:prim}(b) shows the
tight-binding (see Appendix~\ref{sec:tb}) band structure with
($\delta=\pm1$, red lines)
and without ($\delta=0$, blue lines) dimerization.
The direct band gaps at the TRIM points $L$ and $T$ close at
$\delta = 0$  where the parity eigenvalues of the states near
the Fermi level reverse sign by the dimerization sign reversal,
indicating  band inversions at these TRIM points.

{\it Parity--} The number of negative-parity eigenstates
at the TRIM points is listed in
Table \ref{tab:parity} for the two different dimerization states,
$\delta=\pm 1$.
The change of parity states upon dimerization reversal is also
related to the multiple choices of inversion center.
For instance, if one takes \small{$(0,0,0)$}
(the diagonal corner of the primitive cell)
to be the inversion center instead of
\small{$(1/2,1/2,1/2)$}
(the center of the primitive cell),
the parity of the state changes as if the dimerization is reversed.
This is because a structure with reversed
dimerization is equivalent to 
one that is translated by half the cell diagonal.

\begin{table}[b]
	\caption{
		Number of negative parity states $n^-_{\lambda}$
		of the six occupied bands at the TRIM points for
		two dimerizations
		$\delta=\pm1$, classified under the eigenvalues of
		the symmetry operations
		$\sigma^{(1\bar{1}0)}$, $\hat{C}_3^{[111]}$, and
		$\hat{C}_2^{[1\bar{1}0]}$.
		The origin of the parity operation is 	
		\small{$(1/2,1/2,1/2)$}
		and the two-fold $\hat{C}_2$ rotation axis, [1$\bar{1}$0] 
		is normal to the $\sigma$ mirror plane, (1$\bar{1}$0).
		Only those TRIM points  which are invariant under
		these symmetry operations are listed.
		\label{tab:parity}
	}
	\begin{tabularx}{\columnwidth}{
			C{1}C{1}|C{1}|C{1}C{1}|C{1}C{1}C{1}|C{1}C{1} }
		\hline
		\\[-0.8em]
		&
		$\lambda$ & 
		$n^-_{\lambda}$ &
		\multicolumn{2}{c|}{$\sigma^{(1\bar{1}0)}$}
		& \multicolumn{3}{c|}{$\hat{C}_3^{(111)}$}
		& \multicolumn{2}{c}{$\hat{C}_2^{(1\bar{1}0)}$} \\
		$\delta$ & {\scriptsize TRIM} & total & $-i$ & $+i$
		& $-\pi/3$ & $\pi$ & $+\pi/3$
		& $-\pi/2$ & $+\pi/2$ \\
		\hline
		\multirow{4}{*}{$+1$} & $\Gamma$
		    & 0 & 0 & 0 & 0 & 0 & 0 &  0 & 0 \\
		& T & 2 & 1 & 1 & 1 & 0 & 1 &  1 & 1 \\
		& F & 4 & 2 & 2 & - & - & - &  2 & 2 \\
		& L & 2 & 1 & 1 & - & - & - &  1 & 1 \\
		\hline
		\multirow{4}{*}{$-1$} & $\Gamma$
		    & 0 & 0 & 0 & 0 & 0 & 0 & 0 & 0 \\
		& T & 4 & 2 & 2 & 1 & 2 & 1 & 2 & 2 \\
		& F & 4 & 2 & 2 & - & - & - & 2 & 2 \\
		& L & 4 & 2 & 2 & - & - & - & 2 & 2 \\
		\hline
	\end{tabularx}
\end{table}

Topological phases protected by time-reversal or
crystalline symmetries should be independent
of the choice of inversion center as well as the sign of 
dimerization.
Even though there is a parity sign flip at the $L$ and $T$
points, we show below
that the well-known topological
phases of bulk Bi are indeed intact under dimerization
reversal
by calculating the various topological indices: 
(i) $\nu_0$ for STI under time-reversal symmetry,
(ii) $\{\nu^{(\pi)}$, $\nu^{(\pm \pi /3)}\}$ for 
higher order topological insulator (HOTI)
under the three-fold rotational symmetry $\hat{C}_3$, and 
(iii) $\{\nu^{(\pi/2)}$, $\nu^{(-\pi/2)}\}$ for
crystalline topological insulator (CTI) under 
the two-fold rotational symmetry $\hat{C}_2$.

First, the STI $\mathbb{Z}_2$ phase, protected
by time-reversal symmetry, is expressed in terms of the parity
eigenvalues of the occupied states at the TRIM points as,
\begin{eqnarray}
	\nu_0 &=& \frac{1}{2}
	\sum_{\lambda}^{\textrm{TRIM}} n_{\lambda}^{-}
	\,\, \textrm{mod} \,\, 2, \\
	&=& \frac{1}{2} \left( n^-_{\Gamma} +
	n^-_{T} + 3n^-_{F} + 3n^-_{L} \right)
	\,\, \textrm{mod} \,\, 2,
\end{eqnarray}
where $n^-_\lambda$ is the number of occupied states with
negative parity at the TRIM point $\lambda$. For both
dimerized phases, we find $\nu_0 = 0$ corresponding to the
trivial phase that is consistent with previous reports.
\cite{Schindler2018,Bansil_Bi_PNAS}

Next, the HOTI phase of Bi is verified
by grouping occupied states at TRIM points into
$\hat{C}_3$ symmetry subspace according to the
rotation eigenvalues of exp$(i\pi)$ and
exp$(\pm i\pi/3)$.
\cite{Schindler2018}
The fact that each subspace is closed under time-reversal symmetry
allows the $\mathbb{Z}_2$ classification for each subspace.
Among the TRIM points, only $\Gamma$ and $T$ points are
invariant under the $\hat{C}_3$ rotation. On the other hand,
for the remaining ($F_{1,2,3}$ and $L_{1,2,3}$)
TRIM which are not invariant under  $\hat{C}_3$,
one can construct linear combination of these three states
(which transform into each other under three-fold rotation; 
$F_1 \rightarrow F_2 \rightarrow F_3 
	\rightarrow F_1$ as well as $L_i$)
to render them $\hat{C}_3$ eigenstates
(see Ref.\cite{Schindler2018} for more details).
The number of linearly combined states with negative parity
for the two subspaces are
$n^-_{\alpha,\pi} = n^-_\alpha$ and $n^-_{\alpha,\pm\pi/3} = 2 n^-_\alpha = 0$ (mod 2),
where $\alpha \in \{F,L\}$. Thus, the topological invariant
for the two subspaces are given by
\begin{eqnarray}
	\nu^{(\pi)} &=& \frac{1}{2} \left(
	n^-_{\Gamma,\pi} +
	n^-_{T,\pi} +
	n^-_{F} +
	n^-_{L}
	\right) \,\, \textrm{mod} \, 2, \\
	\nu^{(\pm \pi/3)} &=& \frac{1}{2} \left(
	n^-_{\Gamma, \tfrac{\pi}{3}} +
	n^-_{T, \tfrac{\pi}{3}} +
	n^-_{\Gamma,-\tfrac{\pi}{3}} +
	n^-_{T,-\tfrac{\pi}{3}}
	\right) \,\, \textrm{mod} \, 2.
\end{eqnarray}
The dimerization sign reversal changes $n^-_{T,\pi}$ and
$n^-_L$ by two, while $\nu^{(\pi)}$ and $\nu^{(\pm\pi/3)}$ 
do not change under modulo 2.
Hence, we confirm that the 
HOTI phase, $\nu^{(\pi)} = \nu^{(\pm \pi / 3)} = 1$
is intact under the dimerization reversal.

Finally it was predicted that bismuth is also a first-order
CTI protected
by a two-fold rotational symmetry $\hat{C}_2$ around the
[1$\bar{1}$0] axis or its symmetric copies
[01$\bar{1}$] and [$\bar{1}$01].\cite{Bansil_Bi_PNAS}
Similarly, with the classification above, the parity states
can be divided into the $\hat{C}_2$ subspace according to
the symmetry eigenvalues of exp$(i\pi/2)$ and exp$(-i\pi/2)$.
In contrast to the HOTI classification,
the $\hat{C}_2$ subspaces are mapped to each other by
time-reversal symmetry, indicating $\nu^{(\pi/2)}=\nu^{(-\pi/2)}$.
The four TRIM points $\{\Gamma,T,F_1,L_1\}$ are invariant under
$\hat{C}_2$.
The remaining states at the $F_{2,3}$ and $L_{2,3}$
points, which are not invariant under $\hat{C}_2$, can be linearly combined so that they become $\hat{C}_2$ eigenstates. 
The number of negative parity eigenvalues
$n^-_{F,L}$ contributes equally to $\nu^{(\pi/2)}$ and
$\nu^{(-\pi/2)}$ with a weighting factor of one. Thus, the
topological indices are given by
\begin{eqnarray}
	\nonumber
	\nu^{(\pi/2)} = \frac{1}{2} (
	&& n^-_{\Gamma,\tfrac{\pi}{2}} +
	n^-_{T,\tfrac{\pi}{2}} +
	n^-_{F,\tfrac{\pi}{2}} + \\
	&& n^-_{L,\tfrac{\pi}{2}} +
	n^-_{F} + n^-_{L}	)
	\,\, \textrm{mod} \,\, 2,\\
	\nonumber
	\nu^{(-\pi/2)} = \frac{1}{2} (
	&& n^-_{\Gamma,-\tfrac{\pi}{2}} +
	n^-_{T,-\tfrac{\pi}{2}} +
	n^-_{F,-\tfrac{\pi}{2}} + \\
	&& n^-_{L,-\tfrac{\pi}{2}} +
	n^-_{F} + n^-_{L}	)
	 \textrm{mod} \,\, 2.
\end{eqnarray}
Each subspace is found to have a strong topology since
$\nu^{(\pi/2)} = \nu^{(-\pi/2)} = 5$ (mod 2) and
7 (mod 2) for positive ($\delta=+1$) and negative
($\delta=-1$) dimerizations, respectively. Therefore, the
rotational-symmetry-protected CTI phase is well reproduced and
is confirmed to be intact under the dimerization reversal.

It is important to note that in contrast to the topological
phases that are invariant under dimerization sign reversal,
the Zak phase depends on the sign of dimerization
(i.e., choice of the unit cell).
For example,
the $\Gamma$ and $T$ points are projected at the
$\tilde{\Gamma}$ point of the (111) surface BZ
[Fig.~\ref{fig:prim}(c)] where the Zak phase
at $\tilde{\Gamma}$ is determined by the parity eigenvalues,
\begin{eqnarray}
	\frac{1}{\pi}\phi_Z(\tilde{\Gamma}) = \frac{1}{2} 
	(n^-_\Gamma + n^-_T) \,\, \textrm{mod} \,\, 2.
\end{eqnarray}
Here, $\phi_Z(\tilde{\Gamma}) = \pi$ and $0$ for positive
($\delta=+1$) and negative ($\delta=-1$) dimerization,
respectively.
Furthermore, the  $F$ and $L$ points are projected on the other
surface TRIM point $\widetilde{M}$, and the Zak phase at
$\widetilde{M}$,
\begin{eqnarray}
	\frac{1}{\pi}\phi_Z(\widetilde{M}) = \frac{1}{2} 
	(n^-_F + n^-_L) \,\, \textrm{mod} \,\, 2,
\end{eqnarray}
is calculated to be identical to $\phi_Z(\tilde{\Gamma})$
for each dimerized state.
Note that systems with a strong topological order
exhibit different Zak phases at the two surface TRIM points,
exp$[i\phi_Z(\tilde{\Gamma}) + i\phi_Z(\widetilde{M})] = -1$,
or equivalently $\nu_0 = 1$ from Eq. (4).
\cite{FuKaneMele_PRL,FuKane_PRB}
The right panels in Fig.~\ref{fig:prim}(a) show the (111) surface
terminations of the two dimerized states where the surface
with low cleavage energy corresponds to the positive dimerization with
$\pi$ Zak phase.
Surprisingly, the non-trivial phase emerges on the surface
that cuts the weak bonds ($\delta=+1$) rather than the
strong bonds ($\delta=-1$). This is counter-intuitive,
especially when compared to the original 1D SSH model.
	
\begin{figure}[t]
	\centering
	\includegraphics[width=8.6cm]{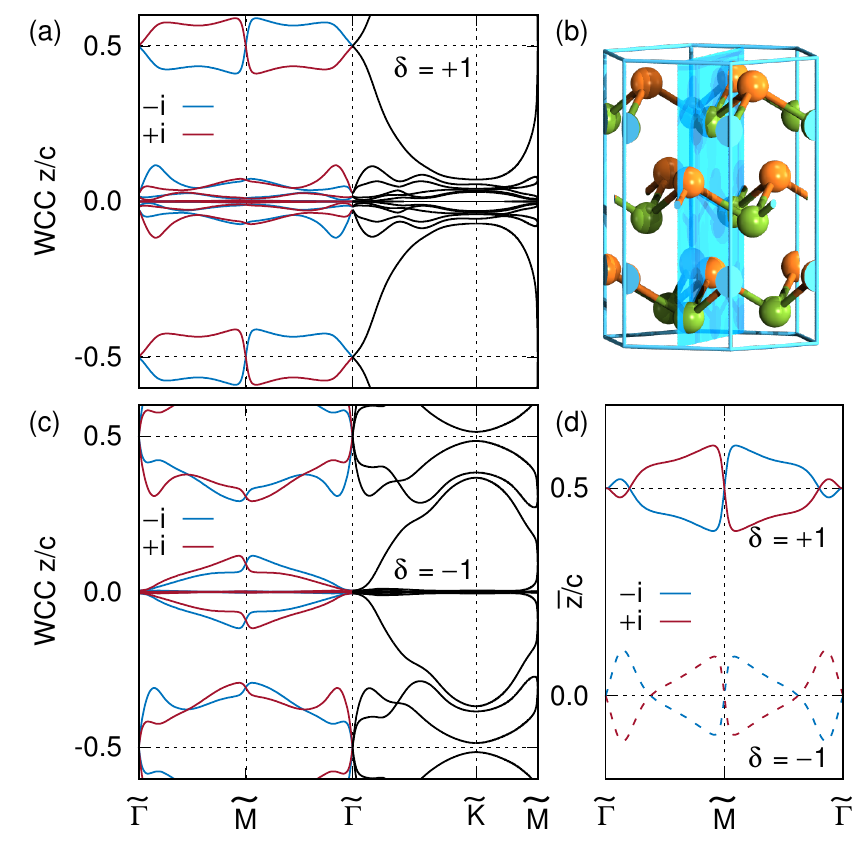}
	\caption{Calculated Wannier charge centers (WCCs) of
		hexagonal Bi for (a) $\delta=+1$ and (c) $\delta=-1$.
		Blue (red) lines denote mirror irreps of $-i$ ($+i$)
		of the mirror plane shown in (b).
		(d) Integrated WCC where solid (dashed) lines correspond to
		$\delta=+1$ ($\delta=-1$).
		\label{fig:hexa_wcc}
	}
\end{figure}

To corroborate the parity analysis,
the hybrid Wannier charge centers (WCCs) are computed for the two
dimerized states in the hexagonal structure having six Bi atoms
(18 valence electrons) as shown in Fig.~\ref{fig:hexa_wcc}.
Because of the inversion and time-reversal symmetries,
the	WCCs are mapped to symmetric copies as
$r_i(\mathbf{k}) \rightarrow -r_i(\mathbf{k})$
and $r_i(\mathbf{k}) \rightarrow r_i(\mathbf{-k})$,
respectively.
For $\delta=+1$, there are two WCCs crossing the cell boundary
$z/c = \pm 0.5$
at $\widetilde{\Gamma}$ ($\widetilde{M}$) which is equal to
the negative parity difference,
$|n^-_\Gamma - n^-_T|=2$ ($|n^-_F - n^-_L|=2$).
Similarly, for $\delta=-1$, four (zero) WCCs cross the cell
boundary at $\widetilde{\Gamma}$ ($\widetilde{M}$), which also
agrees well with the difference of negative parity states.
The factor 2 from the spin degeneracy can be decomposed by grouping
the WCCs based on the mirror eigenvalues $\pm i$ on the
$\widetilde{\Gamma}$-$\widetilde{M}$ plane
(see Fig.\ref{fig:hexa_wcc}).
The WCCs with mirror eigenvalues $-i$ and $+i$ are denoted by blue and red lines, respectively.
It is clearly seen that the number of boundary-crossing 
WCCs in each subspace is reduced by half. For example,
a single blue line passes the boundary in Fig.~\ref{fig:hexa_wcc}(a).
It agrees well with the mirror-symmetry-classified
parity states in Table \ref{tab:parity} where
the negative parity states are divided in half ($n^-_{\lambda,\pm i} = (1/2) n^-_{\lambda}$)
as well as their difference.
Note that the number of negative parity states only provides an
upper limit on the number of WCCs crossing the boundary, with the exact number
of crossings being determined by the symmetry protection.
\cite{Alex_PRB2014}
The net Wannier center $\bar{r} = \sum_i r_i$ is shown in
Fig.~\ref{fig:hexa_wcc}(d), where the half polarization of the
$\pi$ Zak phase $(\delta=+1)$ is clearly seen at the two TRIM, 
consistent with the parity results.

\section{Antiphase Domain Walls in Bi}
\label{sec:dw}

\begin{figure}
	\centering
	\includegraphics[width=8.6cm]{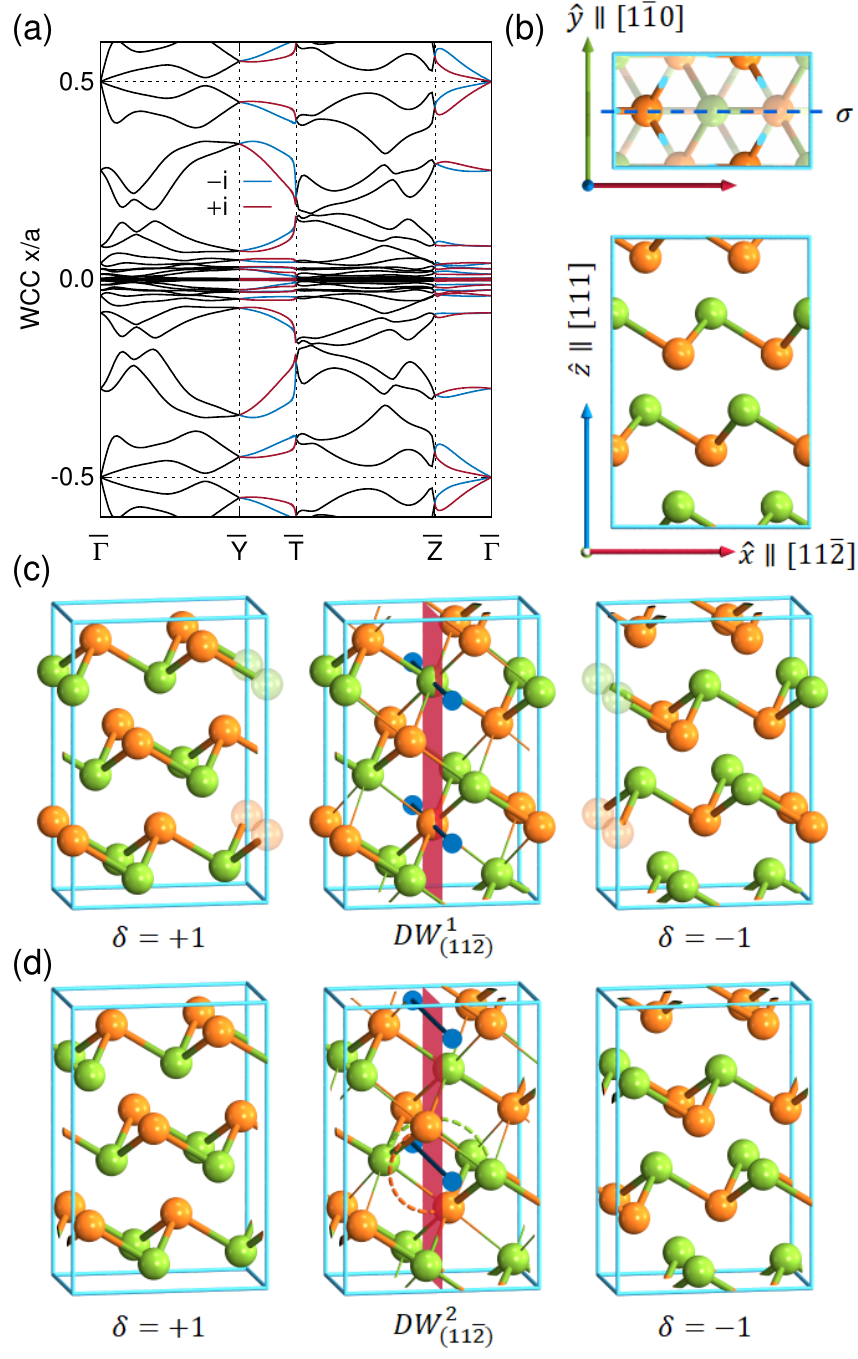}
	\caption{Bi (11$\bar{2}$) domain wall (DW) structure:
	(a) Calculated Wannier charge center (WCC) of the
	slab structure ($\delta=+1$) shown in (c).
	Blue and red lines denote mirror irreps of $-i$ and $+i$, respectively.
	(b) Top-down  and side views of the orthogonal unit cell, where
	$\sigma$ denotes the mirror plane.
	Two types of (11$\bar{2}$) DWs: (c) DW$^1_{(11\bar{2})}$ and (d) DW$^2_{(11\bar{2})}$,
	where the red plane denotes the DW and the blue axes with disks at the end denote the
	rotation or screw axes. The central DW region is sandwiched between two semi-infinite pristine regions with opposite dimerization.
	Type-I DW, DW$^1_{(11\bar{2})}$, passes through the ions and has
	inversion, mirror, and two-fold rotation symmetries. 
	Type-II DW, DW$^2_{(11\bar{2})}$ has the same symmetries but with the two-fold rotation
	replaced with the screw operation, denoted by the dashed curve.
	\label{fig:dw11-2str}
	}
\end{figure}

The corresponding boundary states of the $\pi$ Zak phase
can be realized on the (111) surface by appropriate choice of the
surface termination that is usually hard to control. Fortunately,
Bi is found to exhibit the $\pi$ Zak phase on the 
low-cleavage-energy surface. 
In general, however, the surface with $\pi$ Zak
phase is susceptible to reconstruction and contamination.
\cite{Murakami_PRR_2020}
Thus, instead of the bare surface, we consider antiphase DWs
across which the sign of dimerization is reversed, 
$\delta = \pm 1 \rightarrow \mp 1$, as shown in 
Fig.~\ref{fig:dw111str} and Fig.~\ref{fig:dw11-2str}.
The Bi (111) DWs, hosting the $\pi$ Zak phase,
is indeed the 3D analogue of the 1D SSH model.
The DW is tolerant to chemical contamination and can easily be
found in a system exhibiting charge density wave.

In order to study the DW state without the interference from
the neighbor DW, we use the interface Green's function method
\cite{Sancho_1985,Ozaki_PRB2010}
where the central DW structure is sandwiched between
two semi-infinite pristine Bi with opposite dimerizations,
as is shown in Fig.~\ref{fig:dw111str}(b,c) for the (111) DW and Fig.~\ref{fig:dw11-2str}(c,d) for the (11$\bar2$) DW.
The construction of the Hamiltonian matrices is described 
in Appendix~\ref{sec:H_DW}.
Throughout the remaining manuscript, the tilde ($\sim$) and 
bar ($-$) symbols over the $k$-point labels
denote TRIM points on the (111) and (11$\bar{2}$) DW BZ, 
respectively.

\subsection{Non-trivial (111) Domain Wall}
\label{sec:dw_111}

\begin{figure}[t]
	\centering
	\includegraphics[width=8.6cm]{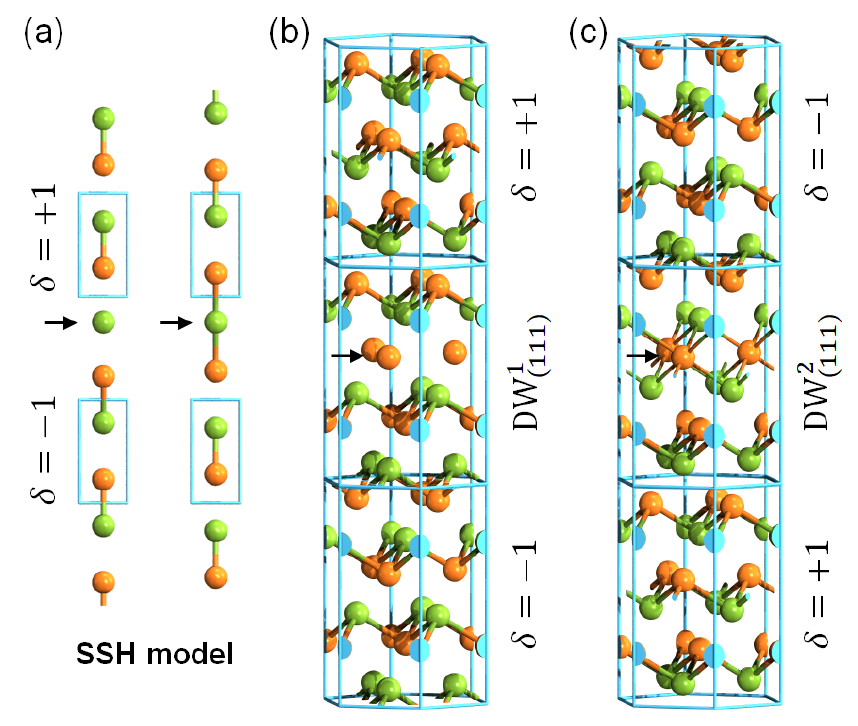}
	\caption{
		(a) Schematic illustration of Su-Schrieffer-Heeger
		(SSH) model with two types of domain-wall (DW)
		interfaces.
		Blue rectangles denotes unit cells and the orange and green
		spheres indicate two sublattice sites.
		(b,c) Two types of Bi (111) DWs: (b) DW$^1_{(111)}$ and (c) DW$^2_{(111)}$, as the 3D analogues
		of the two DWs of the SSH model shown in (a).
		Hexagonal blocks with $\delta = \pm 1$ are the
		conventional cells with either of the dimerizations.
		The central hexagonal block denotes the interface where
		the sign of dimerization flips.
		Inversion symmetry is preserved in both DWs;
		the ion at the inversion center is marked with arrows.
		\label{fig:dw111str}
	}
\end{figure}

\begin{figure}
	\centering
	\includegraphics[width=8.4cm]{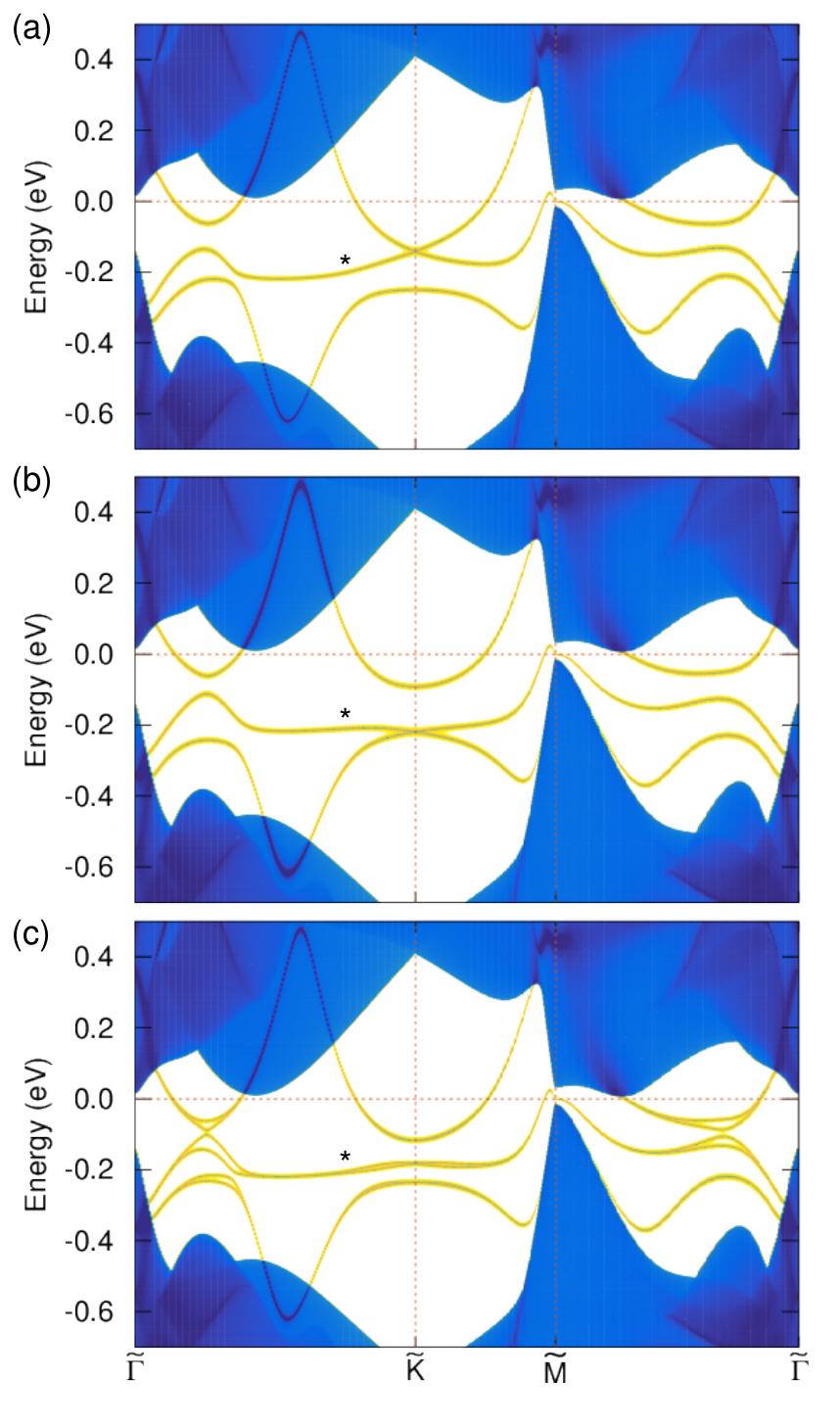}
	\caption{(111) DW band structure calculated using
		the	interface Green's function method for (a) DW$^1_{(111)}$, 
		(b) DW$^2_{(111)}$, and (c) DW$^3_{(111)}$.
		DW-localized states (yellow lines) emerge inside
		DW-projected bulk states (blue shade).	
		The DW localized band 
		labeled with an asterisk is half-filled.
		\label{fig:dw111intf}
	}
\end{figure}

{\it DW localized states --}
We have considered two types of (111) DWs shown in
Fig.~\ref{fig:dw111str}(b,c).
In type-I DW (DW$^1_{(111)}$) the semi-infinite regions below
(above) the DW has $\delta=-1$ ($\delta=+1$) dimerization.
The central DW region has an inversion center, denoted by the
horizontal black arrow, located on an atomic layer
which is weakly bonded with its neighboring atomic layers
along the stacking direction.
In type-II DW (DW$^2_{(111)}$), the sign of dimerization is
opposite and the central layer has strong bondings in both
directions. The DW$^1_{(111)}$ and DW$^2_{(111)}$ correspond
to the two types of DWs of the SSH model shown in 
Fig.~\ref{fig:dw111str}(a),

The calculated DW spectral function is shown in 
Fig.~\ref{fig:dw111intf}(a,b)
where the DW localized states (yellow lines) emerge inside
the DW-projected bulk states (blue shade).
Since inversion symmetry is preserved at the DW, all bands
including the DW-localized yellow bands are doubly degenerate.
Note that, regarding the interface band degeneracy,
the DW localized states resemble Fig.~\ref{fig:schematic_b}(d) instead of Fig.~\ref{fig:schematic_b}(e) even with strong SOC.
This is due to the inversion symmetry at the DW in a sharp
contrast to the bare surface where the inversion symmetry is 
always broken.
The $\pi$ Zak phase at $\tilde{\Gamma}$ and $\widetilde{M}$ 
points induces an odd number of bands inside the
bandgap that guarantees at least one band to be pinned
at the Fermi level as long as the chiral symmetry persists.
We find three DW localized states which are 
buried in the bulk bands at $\tilde{\Gamma}$.
At $\widetilde{M}$, however, the second band in the middle 
among them
appears inside the bandgap indicating adequate chiral symmetry
at the $\widetilde{M}$ point compared to $\tilde{\Gamma}$.
\footnote{
	Our tight-binding parameters are tuned
	to open an insulating gap at, for example, $\tilde{\Gamma}$, 
	that is otherwise closed due to the overlap of electron and 
	hole pockets (refers to Appendix~\ref{sec:tb} for details). 
	The buried DW localized state, therefore, does not worsen 
	the topological property of the DW.
}
As discussed in Sec.~\ref{sec:intro}, the $\pi$ Zak phase corresponds to 
half polarization resulting in $e/2$ modulo $e$ surface charge
per surface unit cell and half-filled in-gap state
\cite{Murakami_PRR_2020}
(i.e., one electron per Kramers' pair
\cite{Zhang_KramersPair}). 
Integration of the spectral function at $\widetilde{M}$
indeed confirms the half-filling of the in-gap state in both 
types of DWs.
The DW localized band, marked with asterisk in Fig.~\ref{fig:dw111intf},
which emerges from the non-trivial state at $\widetilde{M}$
is half-filled and hence metallic.

The number of DW-localized states can also be interpreted 
as the number of bonds truncated at the DW. 
On the (111) surface, a single Bi ion 
per unit cell is exposed with three bonds,
consistent with the number of in-gap states.
The number of bonds truncated at the DW is then determined
by considering the Wannier function center,
which is related to the Zak phase (see Eqs.~\ref{eq:Zak} and \ref{eq:dipole}).
It is noteworthy that, for general systems with complicated
terminations and reduced symmetries, a Green's function approach
can rigorously predict the 
number of surface or interface in-gap states
\cite{Rhim_PRB_2018}
without suffering from the ion-truncating termination
\cite{Murakami_PRX_2018}
or lack of inversion symmetry.
The metallic origin of the DW$^1_{(111)}$
can be simply understood from its construction involving 
the intercalation of a monolayer in pristine bulk Bi, 
which in turn introduces three doubly degenerate bands 
near the Fermi level, where the second band is half filled 
since the number of available electrons is three.

The emergence of 2D Dirac cones at $\tilde{K}$ points in both DWs
is unexpected and the crossing point is found to be lifted upon 
breaking the DW inversion symmetry.
One way to break the inversion symmetry is to vertically translate
the monolayer of DW$^1_{(111)}$. The translation eventually 
leads to a structure, equivalent to the DW$^2_{(111)}$,
having a Bi tri-layer that recovers the inversion symmetry.
Therefore, although the two DW structures 
[Fig.~\ref{fig:dw111str}(b,c)] represent the 
3D analogue of the SSH model, there is a general (111) DW 
structure without the DW inversion symmetry that will be referred 
to as type-III DW, DW$^3_{(111)}$ (see Appendix~\ref{sec:H_DW}
for the DW Hamiltonian).
Figure~\ref{fig:dw111intf}(c) shows the calculated band structure of DW$^3_{(111)}$ where the breakdown 
of inversion symmetry lifts the two-fold
degeneracy of DW localized bands at generic $k$ points
except at the surface TRIM. 
One significant difference of DW$^3_{(111)}$ compared to the 
type-I and type-II DWs, is the splitting of the Dirac crossing 
at $\tilde{K}$, which in turn forms three separate bands, 
indicating that the Dirac cone is related to the
DW inversion symmetry rather than the $\pi$ Zak phase.

\subsection{Trivial (11$\bar2$) Domain Wall}
\label{sec:dw_112}

{\it DW structure --} In this section we consider two types
of ($11\bar2$) DWs as shown in Fig.~\ref{fig:dw11-2str}(c,d)
where the dimerization $[111]$ direction lies on the DW plane.
Thus, the ($11\bar2$) DWs can not be directly compared with the
SSH model, in contrast to the (111) DW 
where its dimerization direction is normal to
the DW plane that is a natural
extension of the 1D SSH model (Fig.~\ref{fig:schematic}). 
Nevertheless, this raises the
question of the emergence of ($11\bar{2}$) DW localized states
and their topological nature.
In both DW types, the central DW region is sandwiched between
two semi-infinite pristine regions with opposite dimerization,
involving a rigid shift of the right semi-infinite region relative
to the left  along the $[111]$ direction by $(c/2)\hat{\bm{z}}$,
or vice versa.
Detailed symmetries of the two DWs and consequent 
degeneracies of the band structure are further discussed in Appendix~\ref{sec:symm_DW}.

\begin{figure}[t]
	\centering
	\includegraphics[width=8.6cm]{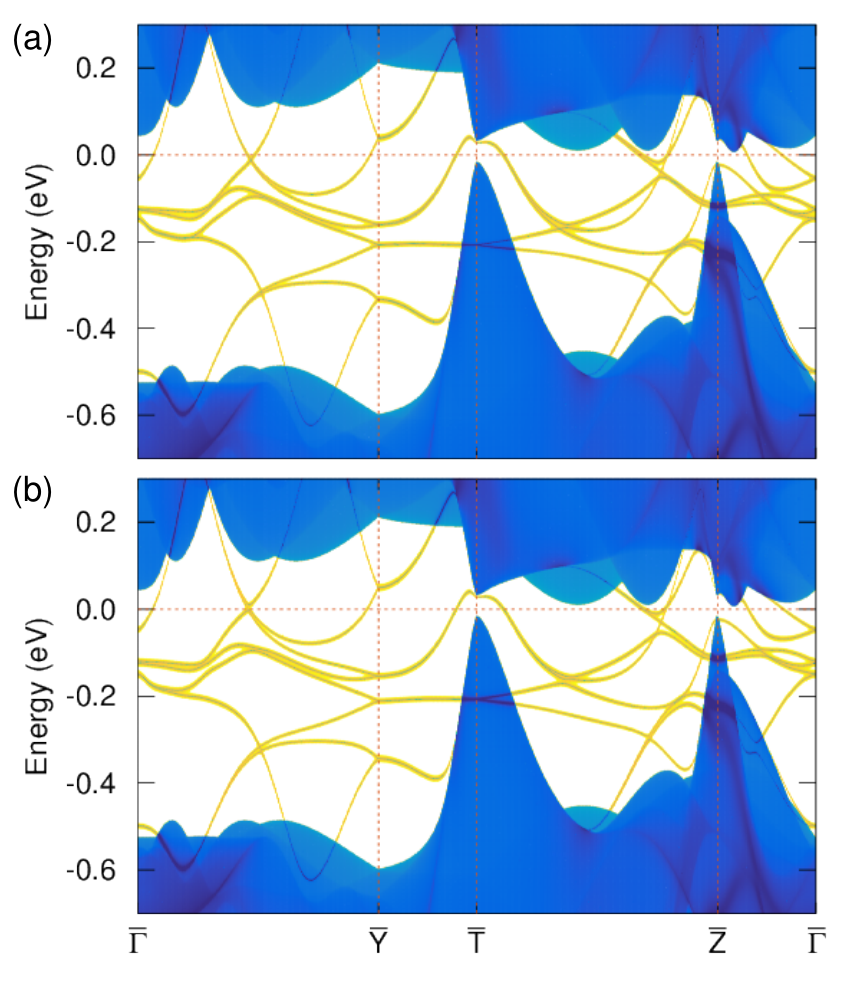}
	\caption{
		($11\bar{2}$) DW band structures calculated using
		the interface Green's function method for
		(a) DW$^1_{(11\bar{2})}$ and (b) DW$^2_{(11\bar{2})}$,
		where the spin-degenerate DW-localized bands (yellow
		lines) emerge in the
		DW-projected bulk bands (blue shade).
		\label{fig:dw11-2intf}
	}
\end{figure}

{\it DW Parity --} 
Figure~\ref{fig:prim}(c) shows the bulk and (11$\bar{2}$) 
interface BZs, where the bulk TRIM points are projected 
on the following interface TRIM points,
\begin{eqnarray}
     \Gamma, F \rightarrow	\bar{\Gamma} ; \hspace{0.1 in}
  F, F  \rightarrow \bar{Y} ;  \hspace{0.1 in}
     T, L  \rightarrow \bar{Z}; \hspace{0.1 in}
     L, L \rightarrow \bar{T}.
\end{eqnarray}
The parity flip induced by the dimerization reversal occurs at
both the $T$ and $L$ points which are projected on the
$\bar{Z}$ and $\bar{T}$ points of the (11$\bar{2}$) interface BZ.
Since the difference in the number of parity flips 
(Table~\ref{tab:parity}) between the
two dimerized domains is zero at $\bar{\Gamma}$ and $\bar{Y}$
and four at $\bar{Z}$ and $\bar{T}$, the DW-projected Zak phases
of both domains are the same, indicating the trivial 
topology of the DW states for both types of (11$\bar2$) DWs,
unless certain crystal symmetry separates each band inversion.
Since the mirror 
plane $\sigma$ in Fig.~\ref{fig:dw11-2str}(b)
is common in both domains and the DWs one can group
the parity states according to the mirror eigenvalues.
Fig.~\ref{fig:dw11-2str}(a) displays the  hybrid WCCs labeled by the
mirror-symmetry
eigenvalues on $\bar{\Gamma}-\bar{Z}$ and $\bar{Y}-\bar{T}$,
which shows no evidence for non-trivial DW state. 

{\it DW localized states and symmetry --}
The band structures of the DW$^1_{(11\bar{2})}$ and
DW$^2_{(11\bar{2})}$ are shown in
Fig.~\ref{fig:dw11-2intf}, where eight of spin-degenerate 
DW-localized bands (yellow lines) appear inside the DW-projected
bulk states (blue shade). 
The number of DW localized bands is related
to the number of truncated bonds on the ($11\bar{2}$) plane
which are eight in both DWs (see red planes in 
	Fig.~\ref{fig:dw11-2str}(c,d)).
The fact that the number of in-gap states is even is
consistent with the trivial Zak phase determined from the product of parity eigenvalues.
Because of the complicated band dispersion and crossings of
the DW localized states, we focus only on the high symmetry line
$\bar{Y} - \bar{T}$ on which the DW localized states 
are approximately four-fold degenerate.
We find no specific crystal symmetry protecting such degeneracy.
Nevertheless, an effective symmetry can be defined
which can give rise to such degeneracy in the thick DW limit (Appendix~\ref{sec:symm_DW}).

\subsection{Arbitrary DW orientation}
\label{sec:dw_arbitrary}

So far, we have considered Bi antiphase DW as a 3D analog of the
SSH model with DW orientation either perpendicular or parallel
to the dimerization direction.
For the (111) DW, the projected parity flips across the DW
inducing the $\pi$ Zak phase while the Zak phase is $0$ for 
the (11$\bar{2}$) DW.
In order to predict the general behavior of the Zak phase
for different DW orientations,
we consider the possible ways of projecting the bulk TRIM points
on various DW planes.
For a surface or DW plane with Miller indices 
$(m_1,m_2,m_3)$,
the surface/interface normal vector is given by,
\begin{eqnarray}
	\label{eq:surface_vector}
	{\bm G}_{\{m_i\}} &=& m_1 {\bm{b}}_1 +  m_2 {\bm b}_2 + m_3 {\bm b}_3,
\end{eqnarray}
where the {\bf b}$_i$'s are reciprocal lattice vectors and
the $\{m_i\} \in \mathbb{Z}$ have no common factor.
A pair of bulk TRIM points
$\{ {\bm \lambda}_{\{n_i\}}, {\bm \lambda}_{\{n^0_i\}} \}$
projected at the same point of the surface/interface BZ are 
always separated by ${\bm G}_{\{m_i\}} /2$, that is given by,
\begin{eqnarray}
	\label{eq:trim}
	{\bm {\lambda}}_{\{n_i\}} &=& \frac{1}{2} (
	n_1 {\bm b}_1 + n_2 {\bm b}_2 + n_3 {\bm b}_3 ), \\	
	{\bm \lambda}_{\{n_i\}} &-& {\bm \lambda}_{\{n^0_i\}} + {\bm G} = \frac{1}{2} {\bm G}_{\{m_i\}}
\end{eqnarray}
where $n_i = \{0,1\}$ selects one TRIM point out of the eight
and $\bf {G}$ is an appropriate reciprocal lattice translation.
The pair of TRIM points
$\{ \bm {\lambda}_n, \bm {\lambda}_{n^0} \}$
satisfy the following relation,
\begin{eqnarray}
	\label{eq:criteria}
	n_i = n_i^0 + m_i - 2 |\bm{G}_i| = (n_i^0 + m_i) \,\, \textrm{mod} \,\, 2.
\end{eqnarray}
This demonstrates that the $n_i$ and $n_i^0$ are identical if the
Miller index $m_i$ is even, otherwise they differ by one if
$m_i$ is odd.
Using this relation, one can enumerate all possible
pairs of TRIM points which overlap on the projected 2D BZ
of an arbitrary surface or DW, which are listed in
Table \ref{tab:projection}.
The (111) and (11$\bar{2}$) DWs correspond to (o,o,o) and
(o,o,e) indices, respectively.
The parity sign flip of Bi induced by dimerization reversal occurs
at $k_2,k_3,k_4$, and $k_8$ points in this notation.
The antiphase DWs with Miller indices (e,e,o), (e,o,e),
and (o,e,e) are expected to have parity sign flip across
the DW giving rise to DW-localized states
similar to the (111) DW or the SSH model.
The remaining (e,o,o) and (o,e,o) DWs are expected to be
trivial similar to the (11$\bar{2}$) DW.
It is important to emphasize that since Table \ref{tab:projection}
is valid for arbitrary reciprocal lattice vectors,
$\bm{b}_i \, (i=1-3)$, it can be applied to a general
centrosymmetric system. The only information required to predict
a non-trivial DW orientation is to determine which TRIM point
flips its parity product across the DW.
It is even easier for bare surfaces, where the parity
eigenvalues of the ground state are enough to predict a 
non-trivial surface orientation.

\begin{table} [t]
	\caption{List of projection of the eight bulk TRIM points
		${\bm {\lambda}}_{\{n_i\}}$,
		each labeled by the set of integers ($n_1,n_2,n_3$), 
		[Eq. (\ref{eq:trim})] 
		on a general
		surface or interface plane with Miller indices ($m_1,m_2,m_3$), 
		labeled as odd (o) or even (e) 
		[see Eq. (\ref{eq:criteria})].
		Also we list the four pairs of TRIM points which overlap on the projected 2D BZ.
		\label{tab:projection}}
	\begin{tabularx}{\columnwidth}{ lC{1} }
		\begin{tabular}{ cc }
			\hline
			\multicolumn{2}{c}{TRIM} \\
			$\bm{\lambda}_{\{n_i\}}$ & $(n_1,n_2,n_3)$ \\
			\hline
			$k_1$ & (0,0,0) \\
			$k_2$ & (1,0,0) \\
			$k_3$ & (0,1,0) \\
			$k_4$ & (0,0,1) \\
			$k_5$ & (0,1,1) \\
			$k_6$ & (1,0,1) \\
			$k_7$ & (1,1,0) \\
			$k_8$ & (1,1,1) \\
			\hline
		\end{tabular} &
		\begin{tabular}{ cc }
			\hline
			Miller indices & \multirow{2}{*}{Pair of TRIM} \\
			$(m_1,m_2,m_3)$ & \\
			\hline
			(e,e,e) & - \\
			(o,e,e) & $\{ k_1k_2,k_3k_7,k_4k_6,k_5k_8 \}$ \\
			(e,o,e) & $\{ k_1k_3,k_2k_7,k_4k_5,k_6k_8 \}$ \\
			(e,e,o) & $\{ k_1k_4,k_2k_6,k_3k_5,k_7k_8 \}$ \\
			(e,o,o) & $\{ k_1k_5,k_6k_7,k_2k_8,k_3k_4 \}$ \\
			(o,e,o) & $\{ k_1k_6,k_5k_7,k_2k_4,k_3k_8 \}$ \\
			(o,o,e) & $\{ k_1k_7,k_5k_6,k_2k_3,k_4k_8 \}$ \\
			(o,o,o) & $\{ k_1k_8,k_2k_5,k_3k_6,k_4k_7 \}$ \\
			\hline
		\end{tabular}
	\end{tabularx}
\end{table}

\subsection{Experimental Realization of Dimerization Reversal via Optical Pumping }
\label{sec:exp_realization}

There are two plausible experimental approaches to realize Bi
antiphase DWs. The first approach is to search for dislocation
defects in a Bi single crystal.
For example, the (11$\bar{2}$) DW would appear on the
(111) surface as a half step edge (step height of 
Bi monolayer, $c/6$) in
scanning tunneling microscopy measurements.
The second approach is to induce local dimerization reversal
in pristine Bi using intense femtosecond laser-pump excitations,
which have shown the reduction of the equilibrium displacement
($\Delta_0$) of Bi, referred to as ``ultrafast bond softening".
\cite{Fritz_Science}
More specifically, the laser-pump promotes valence electron into
the conduction band and softens the Bi bond that agrees well with
complementary density functional theory calculations.
\cite{Fritz_Science}
The calculations also predict a transient structural transition to
undimerized state ($\Delta_0 \rightarrow 0$) upon excitation
of $\sim$2.5\% of valence electrons.
The energy barrier between the two dimerized ground states was found to decrease with increasing charge excitations, thus 
supporting the plausibility of dimerization reversal by excitations.
Indeed, experiments confirmed that excitations higher than 2\%
lead to an irreversible ``damage" to the samples
suggesting that a permanent dimerization reversal may be achieved via the laser-pump excitations.
\cite{Fritz_Science}

\section{Conclusion}
\label{sec:conclusion}

We propose that the $\alpha$-phase of bulk Bi is a 3D
manifestation of the SSH model. We demonstrate that while the
HOTI and CTI phases of bulk Bi remain invariant under dimerization
sign reversal, the Zak phase undergoes a transition from
$\pi$ to 0.
The (111) antiphase DW is found to host metallic DW bands,
which are topologically protected due to the difference in
polarization between the two oppositely dimerized domains
(i.e., $\pi$ Zak phase), which is the 3D analogue of the SSH model.
Although the (11$\bar{2}$) DW has no such polarization difference,
the DW localized states exhibit interesting behavior
related to an effective symmetry that reveals itself in thicker DWs.

To the best of our knowledge, this result is the first
demonstration of the non-trivial Zak phase in 3D antiphase DWs.
Unlike the bare surface being vulnerable to doping, contamination,
or reconstruction, antiphase DWs offer a relatively stable
platform for the manifestation of a non-trivial Zak phase.
Furthermore, the common presence
of DWs in charge-density-wave states offers a novel venue for
investigating the potential of the non-trivial Zak phase.

\section{Methodology}
\label{sec:methods}
The tight-binding parameters are extracted from the Wannier
Hamiltonian obtained by using \small{VASP}-\small{Wannier90}
interface.\cite{Kresse96a,Kresse96b,Mostofi}
The pseudopotentials are of the projector-augmented-wave
type as  implemented in \small{VASP},\cite{Blochl94,KressePAW}
with valence configurations 6$s^2$6$p^3$ for Bi.
The exchange-correlation functional is described by
the Perdew-Burke-Ernzerhof generalized gradient
approximation (PBE). \cite{PBE}
The plane-wave cut-off energy is set to $300$\,eV and
the Brillouin zone sampling grid is
12\,$\times$\,12\,$\times$12.
The structure is relaxed with a constraint of being FCC
for an insulating band gap.
The twelve strongest hopping terms are then used in the
calculation together with atomic spin-orbit coupling
for the $p$ orbitals.
The spectral density of DW-localized states is calculated
using the interface Green's function method
\cite{Sancho_1985,Ozaki_PRB2010}
where two
semi-infinite surface Green's functions are first calculated
for the two dimerized phases and then combined
with the central DW Hamiltonian.

\begin{acknowledgments}
This work at CSUN is supported by the NSF-Partnership in
Research and Education in Materials (PREM) Grant No. DMR-1205734.
D.V.~was supported by NSF Grant DMR-1954856.
\end{acknowledgments}

\appendix

\section{Tight-binding parameters}
\label{sec:tb}

\begin{figure}[b]
	\centering
	\includegraphics[width=8.6cm]{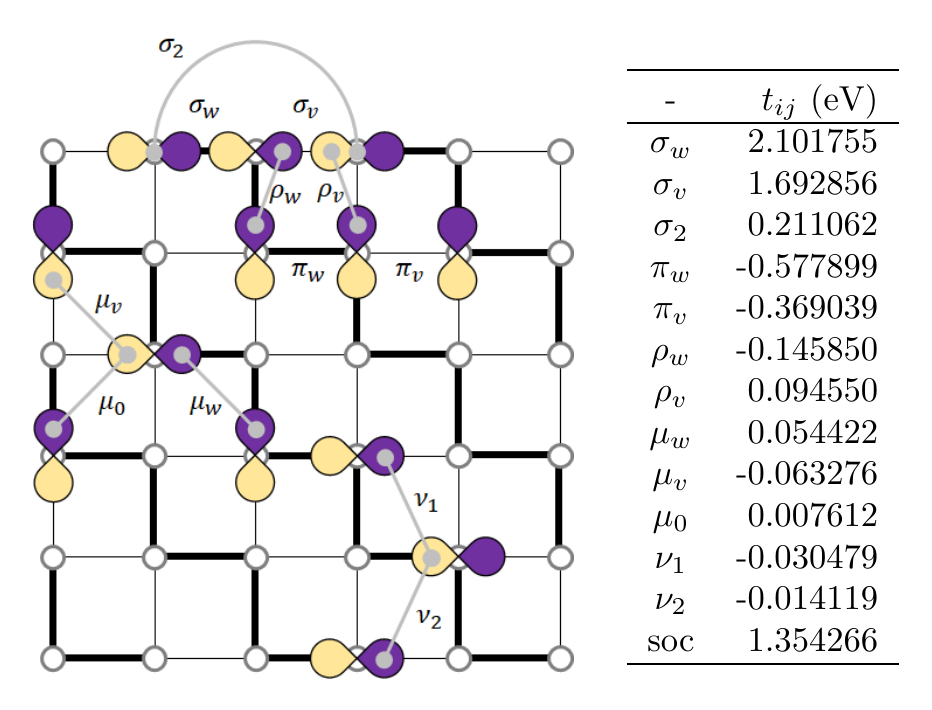}
	\caption{Selected tight-binding parameters for Bi.
		The background square lattice illustrates the Bi plane
		normal to one of the Cartesian coordinate vectors.
		Thick (thin) black line denotes strong (weak)
		bonding after the atomic displacement (i.e. dimerization).
		The hopping terms with subscripts, $w$ and $v$, are lifted
		by the dimerization and they are swapped by the
		dimerization reversal.
		Terms with numerical subscripts are not affected by the
		dimerization reversal.
		\label{fig:parameter}
	}
\end{figure}

Although Bi has finite direct band gap in the whole BZ,
its indirect gap between $T$ and $L$ points is negative
causing metallic band structure and difficulties in the
analysis of topological properties. For instance, the projected
bulk states (blue shade in Fig.~\ref{fig:dw11-2intf}) at the $\bar{Z}$ point on the
(11$\bar{2}$) BZ, should be gapless due to the negative indirect
gap. In order to suppress the complexity, we have constructed the
tight-binding parameters in the fcc instead of the rhombohedral
cell, which in turn opens up a gap, shown in Fig.~\ref{fig:prim}(b),
without affecting the topological properties such as parities
at the TRIM points.

Figure \ref{fig:parameter} shows the selected twelve hopping
terms in a dimerized cubic lattice and the amplitudes are listed
in the inset table together with that of the atomic spin-orbit
coupling.
The $\sigma_{w,v}$, $\pi_{w,v}$, and $\rho_{w,v}$ are the
nearest-neighbor hoppings distinguished by the relative direction
of the $p$ orbitals; the $\sigma_2$ is the third nearest-neighbor
$\sigma$-bond like hopping term;
the $\rho_{w,v}$ terms vanish without dimerization
because of the basis symmetry; the $\mu_{w,v,0}$ and $\nu_{1,2}$
terms are the second nearest-neighbor hoppings, and the
$\nu_{1,2}$ terms do not change under dimerization reversal
unless the direction of dimerization changes.

\section{Hamiltonian of the DW}
\label{sec:H_DW}

\begin{figure}[b]
	\centering
	\includegraphics[width=7.8cm]{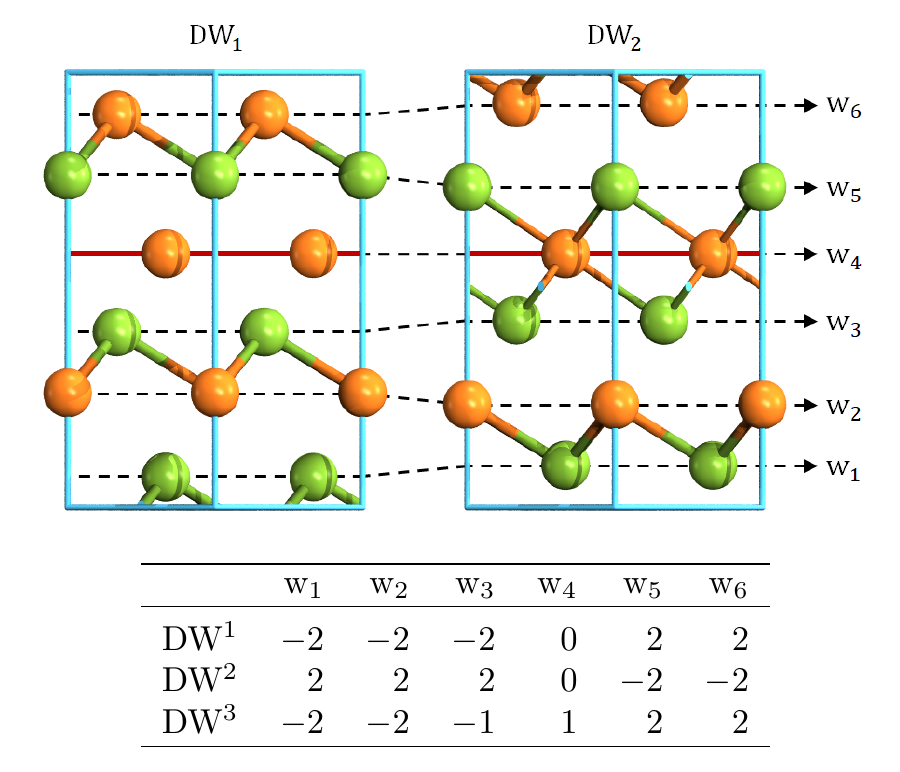}
	\caption{Modulation of hopping parameters across the
		(111) DW. Two types of DWs are illustrated
		together with horizontal planes, separated from the DW
		denoted as red solid line.
		Weight factors used for the linear interpolation are
		presented in the inset table
		together with the type-III DW.
		\label{fig:dw_H1}
	}
\end{figure}

\begin{figure}[h]
	\centering
	\includegraphics[width=7.8cm]{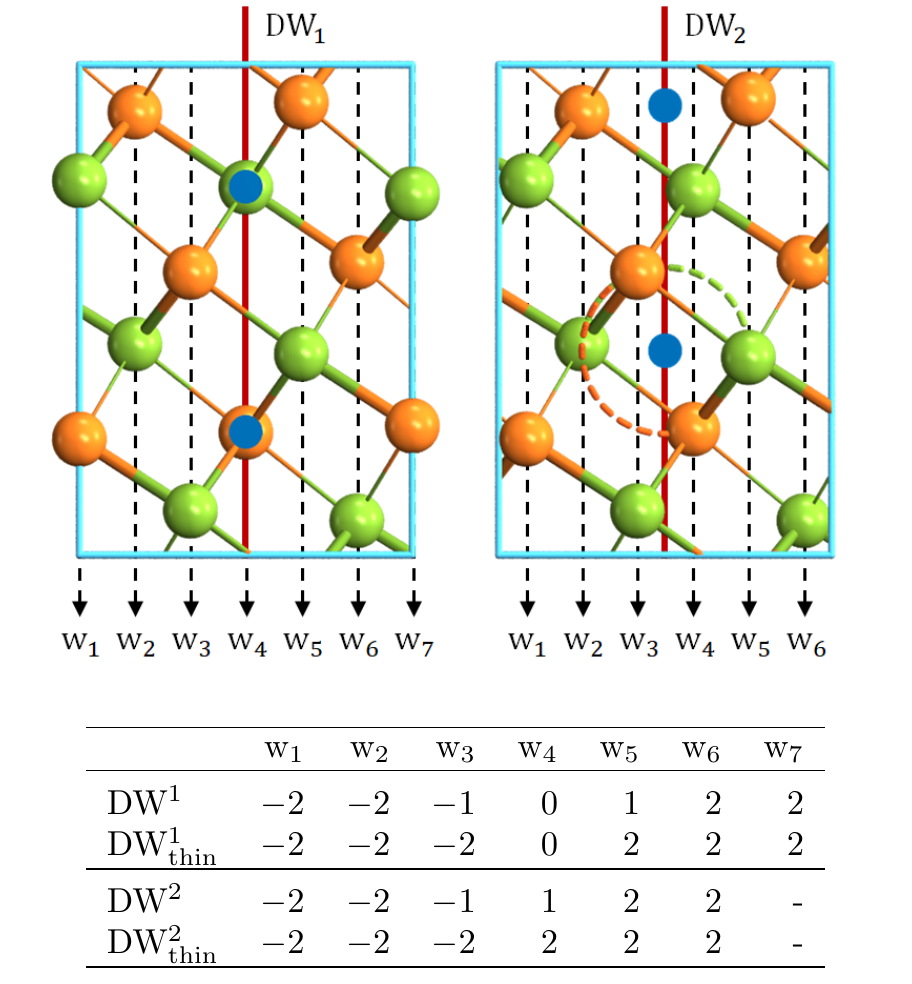}
	\caption{Modulation of hopping parameters across the
		(11$\bar{2}$) DW. Two types of DWs are illustrated
		together with vertical planes, separated from the DW
		denoted as red solid line.
		Weight factors used for the linear interpolation are
		presented in the inset table.
		\label{fig:dw_H}}
\end{figure}

The hopping terms with subscripts $w$ and $v$ modulate in the
vicinity of the DW. The amplitude of the hopping terms is
determined via linear interpolation of the two hopping terms
by considering the distances of two basis from the DW plane.
Namely,
\begin{eqnarray}
	t_{ij}^{'w} &=& (1-r_{ij}) t_{ij}^{w} + r_{ij} t_{ij}^{v}, \\
	t_{ij}^{'v} &=& (1-r_{ij}) t_{ij}^{v} + r_{ij} t_{ij}^{w}, \\
	r_{ij} &=& (w_i + w_j + 4) / 8,
\end{eqnarray}
where $t_{ij}$ is the original hopping terms of Bi
(Fig.~\ref{fig:parameter}) and
$r_{ij}$ is the mixing ratio depending on the weight factor
$w_i$ representing the sign of dimerization
as illustrated in Fig.~\ref{fig:dw_H1} and \ref{fig:dw_H}.
The DW$_{(111)}^1$ and DW$_{(111)}^2$ have 6 atomic layers
along the stacking direction with one ion per layer.
The DW$_{(11\bar{2})}^1$ has 14 ions
and 7 vertical planes ($w_1, \cdots, w_7$) in the cell while
DW$_{(11\bar{2})}^2$ has 12 ions and 6 vertical planes.
In this interpolation scheme, the weighting factors for the
thinnest (11$\bar{2}$) DW are also listed with a subscript 
``thin" in the inset of Fig.~\ref{fig:dw_H}.
The results for thin DW case shown in
Fig.\ref{fig:dw11-2intf_symm}(b,d) are calculated using
Hamiltonians generated with these weighting factors.

\section{Symmetry of $(11\bar{2})$ DW}
\label{sec:symm_DW}

\begin{figure}[t]
	\centering
	\includegraphics[width=8.0cm]{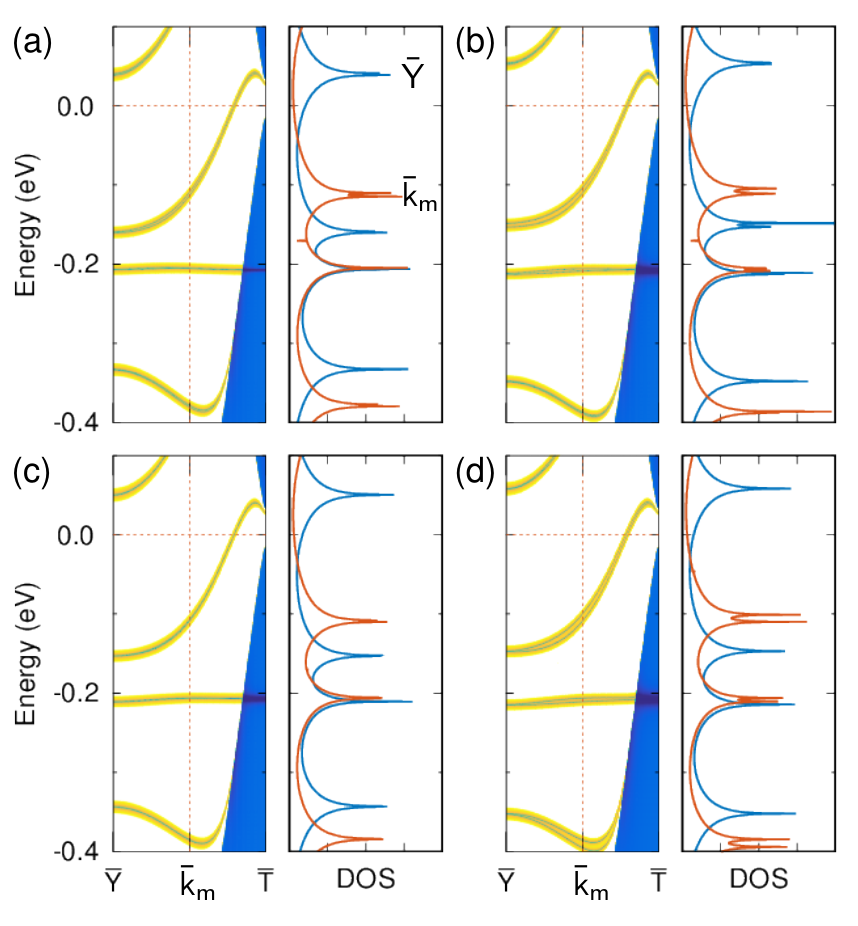}
	\caption{Zoom-in band structure (left panel) and 
		k-resolved spectral function (right panel) 
		for (a,b) DW$^1_{(11\bar{2})}$ 
		and (c,d) DW$^2_{(11\bar{2})}$.
		The spectral functions at $\bar{Y}$ (blue line) 
		and $\bar{k}_m$ (red line) points are plotted 
		in log scale.
		(b) and (d) are similar plots for thinner DW width than 
		(a) and (c), respectively,
		showing the splitting of the peaks.
		\label{fig:dw11-2intf_symm}
	}
\end{figure}

In type-I DW, DW$^1_{(11\bar{2})}$, 
[Fig.~\ref{fig:dw11-2str}(c)] where the Bi atoms
lie on the DW plane, has inversion, mirror, and 
two-fold rotation symmetries. The two-fold rotation, $\hat{C}_2$
around the [1$\bar1$0] direction is denoted by the horizontal blue
axis.
On the other hand, type-II DW, DW$^2_{(11\bar{2})}$, 
[Fig.~\ref{fig:dw11-2str}(d)]
intersects the bonds between atoms across the DW and has similar
symmetries as  DW$^1_{(11\bar{2})}$ except that the $\hat{C}_2$
rotation is replaced by a screw $\hat{S}_2$ symmetry involving
a two-fold rotation around the [1$\bar1$0] direction followed by a
half a translation along the same axis. Namely,
\begin{eqnarray}
	\hat{C}_2 &:& (x,y,z) \rightarrow (-x,y,-z)
	\otimes i \sigma_y, \\
	\hat{S}_2 &:& (x,y,z) \rightarrow (-x,y+1/2,-z)
	\otimes i \sigma_y.
\end{eqnarray}
Because of the DW inversion symmetry,
all bands are two-fold degenerate in the whole interface BZ.
The high symmetry lines along $k_y$
($\bar{\Gamma} - \bar{Y}$ and $\bar{Z} - \bar{T}$)
are invariant under 
the $\hat{C}_2$ operation for the DW$^1_{(11\bar{2})}$ and
the $\hat{S}_2$ operation for the DW$^2_{(11\bar{2})}$.
In addition, the nonsymmorphic $\hat{S}_2$ symmetry for
DW$^2_{(11\bar{2})}$ guarantees a four-fold
degeneracy at $\bar{Y}$ and $\bar{T}$  where
$k_y = \pm \pi$.
\cite{YoungKane_PRL}

In Fig.~\ref{fig:dw11-2intf_symm}(a,c) we display the zoom-in band structure of Fig.~\ref{fig:dw11-2intf}(a,b)
on the high symmetry line $\bar{Y} - \bar{T}$ 
for both types of DWs
along with the corresponding $k$-resolved spectral function (blue
lines) at $\bar{Y}$.
The calculations of the spectral function for DW$^2_{(11\bar{2})}$ corroborate the
emergence of single peaks at $\bar{Y}$ which are indeed four-fold
degenerate.
On the other hand, such a four-fold degeneracy is not protected
by $\hat{C}_2$ symmetry for  DW$^1_{(11\bar{2})}$, which, however,
exhibits similar band folding at $\bar{Y}$ point, where the peaks
in Fig.~\ref{fig:dw11-2intf_symm}(a) have negligible splitting.
Furthermore, the high symmetry line $\bar{Y} - \bar{T}$ appears
to be four fold degenerate in both types of DWs, which cannot
not be explained by the crystal symmetries.
This apparent four-fold degeneracy of the high symmetry line
is found to be lifted as the DW thickness is reduced, as is
clearly shown by the splitting of the peaks (denoted by red) in
Fig.~\ref{fig:dw11-2intf_symm}(b,d).
This implies an effective symmetry which appears to be present
only for thicker DWs.

It is worth to emphasize that the two types of DWs are
distinguished only by the position of DW plane and the
Hamiltonian difference between the two DWs becomes subtle
with increasing DW thickness.
Both Hamiltonians eventually acquire $\hat{C}_2$ and $\hat{S}_2$
symmetries in the thick DW limit. The two symmetries are combined 
to an effective symmetry of the DW which can be expressed as
\begin{eqnarray}
	\hat{C}_2 \hat{S}_2 : (x,y,z) \rightarrow
	(x,y+1/2,z) \otimes -1,
\end{eqnarray}
consisting of a half translation operation 
that allows the BZ-unfolding, $k_y$: [$-\pi$;$+\pi$] 
$\rightarrow$ [$-2\pi$;$+2\pi$] and causes band degeneracy
on the $k_y = \pm \pi$ line. This emergent half translation 
symmetry naturally explains the apparent four-fold degeneracy
(i) of the high symmetry line $\bar{Y} - \bar{T}$ 
in both types of DWs, and 
(ii) at $\bar{Y}$ in DW$^1_{(11\bar{2})}$.

\bibliography{bi_dw}

\end{document}